==================

\documentstyle[12pt]{article}

\begin{document}

\def\lsim{\mathrel{\lower2.5pt\vbox{\lineskip=0pt\baselineskip=0pt
           \hbox{$<$}\hbox{$\sim$}}}}
\def\gsim{\mathrel{\lower2.5pt\vbox{\lineskip=0pt\baselineskip=0pt
           \hbox{$>$}\hbox{$\sim$}}}}

\begin{titlepage}

\title{{\bf Dark Matter with Variable Masses}
\thanks{Work partially supported by CICYT under
contract AEN90--0139.}}

\author{
{\bf Juan Garc\'{\i}a--Bellido}\thanks{Supported by
MEC--FPI Grant. e--mail: bellido@cc.csic.es} \\
Instituto de Estructura de la Materia , \ CSIC\\
Serrano, 123\ \ \ E--28006\ \ Madrid ,\ \ Spain  }

\date{}
\maketitle
\def\baselinestretch{1.15}
\begin{abstract}
String effective theories contain a dilaton scalar field which
couples to gravity, matter and radiation. In general, particle
masses will have different dilaton couplings. We can always
choose a conformal frame in which baryons have constant masses
while (non--baryonic) dark matter have variable masses, in the
context of a scalar--tensor gravity theory. We are
interested in the phenomenology of this scenario. Dark matter
with variable masses could have a measurable effect on the
dynamical motion of the halo of spiral galaxies, which may
affect cold dark matter models of galaxy formation. As a
consequence of variable masses, the energy--momentum tensor is
not conserved; there is a dissipative effect, due to the
dilaton coupling, associated with a ``dark entropy" production.
In particular, if axions had variable masses they could be
diluted away, thus opening the ``axion window". Assuming that
dark matter with variable masses dominates the cosmological
evolution during the matter era, it will affect the primordial
nucleosynthesis predictions on the abundances of light elements.
Furthermore, the dilaton also couples to radiation in the form
of a variable gauge coupling. Experimental bounds will constrain
the parameters of this model.
\end{abstract}

\vskip2.5cm
\begin{flushleft}
Presented at NEUTRINO-92 International Conference on Neutrino
Physics and Astrophysics, Granada, June 1992. Based on work done
with Alberto Casas and Mariano Quiros.
\end{flushleft}

\vskip-22.5cm
\rightline{{\bf IEM--FT--54/92}}
\rightline{{\bf March 1992}}
\vskip3in

\end{titlepage}

\newpage
\def\baselinestretch{1.25}

\section{Introduction}

Most particle physicists believe that the theory of gravity at
low energies (general relativity, scalar-tensor theories, etc.)
is an effective approximation of some fundamental theory of
quantum gravity at energies beyond the Planck scale ($M_{Pl}\sim
10^{19}$ GeV). Nowadays, the only reliable candidates for such a
fundamental theory are superstrings \cite{GSW}. String theory
assumes that elementary particles are extended one--dimensional
objects, as an alternative to the point--like description of
quantum field theory. This simple assumption has very
interesting consequences. In particular, in the low energy limit
of strings moving in curved backgrounds we recover
a generally covariant field theory. However, the string
effective gravity theory is not precisely general
relativity. In fact, all closed bosonic strings contain in their
massless gravitational sector, apart from the graviton, a
dilaton and an antisymmetric tensor. These fields will appear in
the low energy effective theory from strings.

We can write the tree level heterotic string effective action in
four dimensions, keeping only linear terms in the string tension
$\alpha'$ and in the curvature $R$, as \cite{GSW}
\begin{equation}
\label{SEA}
S=\frac{1}{\alpha'} \int d^{4}x \sqrt{-G} e^{-2\phi}
\left(R + 4G^{\mu\nu}\partial_\mu \phi
\partial_\nu \phi - \frac{1}{12} H_{\mu\nu\lambda}
H^{\mu\nu\lambda} + \alpha'{\cal L}_m \right)
\end{equation}
\begin{equation}
\label{SLM}
{\cal L}_m = - \frac{1}{4} F_{\mu\nu}F^{\mu\nu}
- \frac{1}{2} \sum_i \left( \partial_\mu C_i \partial^\mu C_i
+ m_i^2(\phi) C_i^2 \right)
\end{equation}
where $H_{\mu\nu\lambda}$ is the field strength of the
antisymmetric tensor, $F_{\mu\nu}$ is the electromagnetic tensor
and $C_i$ are the matter fields that appear in the
compactification to four dimensions. The dilaton dependence of
$m_i(\phi)$ will be assumed to be linear in the exponential
\begin{equation}
\label{MPH}
m_i(\phi) = m_i\ e^{\gamma_i \phi}
\end{equation}
where $\gamma_i$ parametrizes our ignorance on the details of
supersymmetry breaking in string theory and the low energy
non--perturbative effects. In general, different particles will
have different $\phi$--dependences \cite{CQG}. We will study the
phenomenological signatures of this effective action, the most
characteristic property being the coupling of the dilaton to the
gravitational, gauge and matter sectors. A conformal
redefinition ($G_{\mu\nu} = e^{2\phi} \bar{g}_{\mu\nu}$) allows
us to rewrite the action (\ref{SEA}) in the Einstein frame as
\footnote{We have redefined the scalar field $\phi$ as twice the
dilaton, for convenience, and considered only the matter fields
as classical point particles, disregarding the antisymmetric
field. From here on we will work in
units of Planck's mass, $\alpha'=16\pi$.}
\begin{equation}
\label{SMD}
S = \int d^4x \sqrt{-\bar{g}} \left(\bar{R}
- \frac{1}{2} (\partial\phi)^2
+ 16\pi e^{\beta_I\phi} {\cal L}_{m_I}
+ 16\pi e^{\beta_V\phi} {\cal L}_{m_V}  \right)
\end{equation}
\begin{equation}
\label{LM}
{\cal L}_m =  \frac{-1}{\sqrt{-\bar{g}}} \sum_n\ m_n
\int d\tau_n \left( -\bar{g}_{\mu\nu}(x_n)
\frac{dx^{\mu}_n}{d\tau_n}\frac{dx^{\nu}_n}{d\tau_n}
\right)^{1/2} \delta^{(4)}(x-x_n) ,
\end{equation}
where $n$ labels a set of classical point particles
with variable masses $m_n(\phi) = e^{\beta\phi} m_n$ and
we have assumed only {\em two} different dilaton couplings
$(\beta_I,\beta_V)$ associated with baryonic (visible) and
dark (invisible) matter sectors \footnote{We restrict ourselves,
for simplicity, to the case $\ \beta_I, \beta_V>0$.}. Such a
theory has been considered previously in refs.\cite{DGG,CQG} as
a generalization of Jordan--Brans--Dicke theory of gravity
\cite{JBD,SCT}. It has also been studied in the context of extended
inflation in Ref.\cite{EI}. It explicitly violates the weak
equivalence principle but is not ruled out by experiment.
In fact, the weak equivalence principle has been tested only
with ordinary (baryonic and leptonic) matter and energy
\cite{SCT}, but we know that most of the matter in the universe
is dark matter \cite{KT}. String theory gives no prediction on
$(\beta_I,\beta_V)$ but suggests that this scenario may arise.
We will study the physical consequences of this assumption and
try to obtain as much phenomenological constraints as possible
on the parameters of the model.

We know that physics cannot distinguish between conformal
frames, therefore we are free to choose whatever masses are made
constant. We choose constant visible masses for
convenience since they give constant units of measure and also
visible particles follow trajectories which are geodesics of the
metric. We thus perform a conformal redefinition
\begin{equation}
\label{DEF}
\bar{g}_{\mu\nu} = e^{-2\beta_V \phi} g_{\mu\nu} =
\Phi\ g_{\mu\nu}
\end{equation}
in order to leave constant masses for visible matter.
The resulting theory has the form of a generalized
Jordan--Brans--Dicke theory with variable masses for the dark
matter sector \cite{CQG}
\begin{equation}
\label{SBD}
S = \int d^4x \sqrt{-g} \left(\Phi R -
\frac{\omega}{\Phi} (\partial\Phi)^2
+ 16\pi\Phi^{\sigma}{\cal L}_{m_I}
+ 16\pi{\cal L}_{m_V} \right)
\end{equation}
where the two parameters $(\omega,\sigma)$ are given by
\begin{equation}
\label{OSB}
\begin{array}{rl}
2\omega+3&=
{\displaystyle
\frac{1}{4\beta^2_V} }\vspace{2mm}\\
1-2\sigma&=
{\displaystyle
\frac{\beta_I}{\beta_V} }  \ \ .
\end{array}
\end{equation}

We know that baryonic matter gives a very small contribution to
the critical density of the universe \cite{KT}. We will assume
that the cosmological evolution during the matter era is dominated
by dark matter with {\em variable} masses. The gravitational
equations of motion are then given by \cite{NPB}
\begin{equation}
\label{SEQ1}
R_{\mu \nu}
=\frac{8\pi}{\Phi} \left( \frac{1}{2}g_{\mu \nu}
T^\lambda_{\ \lambda} - T_{\mu \nu} \right) -\frac{\omega}
{\Phi^2} \partial_{\mu}\Phi
\partial_{\nu}\Phi-\frac{1}{\Phi}\left(D_{\mu}D_{\nu}
\Phi+\frac{1}{2}g_{\mu \nu} D^2 \Phi \right)
\end{equation}
\begin{equation}
\label{SEQ2}
(2\omega+3) D^2\Phi=8\pi (1-2\sigma)
T^\lambda_{\ \lambda}
\end{equation}
with the energy--momentum conservation equation
\begin{equation}
\label{EMC}
T^{\mu\nu}_{\ \ ;\nu} = \sigma \frac{\partial^{\mu}\Phi}{\Phi}
\ T^\lambda_{\ \lambda} .
\end{equation}
Note that during the radiation era the energy--momentum tensor
is exactly conserved. In fact, in that era, the scalar field is
constant, as we can see from eq.(\ref{SEQ2}).
The covariant non--conservation of the energy--momentum tensor
just expresses the fact that there is an energy exchange between
dark matter and the scalar, which gives a dissipative effect.

The particle trajectories corresponding to dark matter
are given by
\begin{equation}
\label{GEO}
\frac{d^2x^{\mu}}{d\tau^2} + \Gamma^{\mu}_{\nu\lambda}
\frac{dx^{\nu}}{d\tau}\frac{dx^{\lambda}}{d\tau} +
\sigma \frac{\partial^{\mu}\Phi}{\Phi} = 0
\end{equation}
and therefore do not follow geodesics, being modified by the term
$\ \sigma {\displaystyle \frac{\partial^\mu \Phi}{\Phi}\  }$.
Note that this equation exactly corresponds
to the geodesic of the metric $\ \tilde{g}_{\mu\nu}(x) =
\Phi(x)^{2\sigma} g_{\mu\nu}(x)\ $
since $\ \tilde{\Gamma}^{\mu}_{\nu\lambda}
=\Gamma^{\mu}_{\nu\lambda}-\frac{1}{2}
\partial^\mu (\ln\Phi^{2\sigma})\ g_{\nu\lambda}\ $.
Of course, baryonic particles {\em do} follow geodesics of
$g_{\mu\nu}(x)$. To understand the physical significance of
eq.(\ref{GEO}), let us consider the acceleration of a non
relativistic particle in a weak and static gravitational field
($g_{oo}\simeq -1+\frac{2GM}{r},\; g_{ij}\simeq \delta_{ij},\;
g_{oi}\simeq 0$) \cite{WEI}
\begin{equation}
\label{ACN}
\frac{d^2x^i}{dt^2} + \sigma \frac{\dot{\Phi}}{\Phi}
\frac{dx^i}{dt} \simeq - \frac{GM}{r^2}\frac{x^i}{r},
\end{equation}
which gives the Newtonian acceleration due to the
gravitational attraction of a mass $M$ plus a friction force due
to the variation of mass. Eq.(\ref{ACN}) exactly corresponds
to Newton's second law ${\displaystyle \frac{d}{dt} (m v^i) =
F^i}$, as one would expect from variable masses.

We will try to extract new phenomenological signals of this
theory of gravity with matter coupled to a scalar. As we will
show in the next section, the fact that dark matter with
variable masses do not follow geodesics will affect the
dynamical motion of the halo of spiral galaxies. At scales
of superclusters this effect could change the general
picture of dark matter halos. On the other hand, the
dissipative effect due to the scalar coupling could also give an
important entropy production. If axions had variable masses
this effect can be used to dilute their contribution to the
critical energy density of the universe, therefore eluding the
cosmological bounds on their mass.

\section{Dark matter halo of spiral galaxies}

{}From dynamical observations of spiral galaxies we know that
there is a halo with great amounts of dark matter
\cite{KT,RST}. If the dark matter in the halo were composed of
particles with variable masses, it would have, in principle, a
measurable effect on the dynamical motion of the halo.
The measurements of the halo mass are obtained
from Kepler's third law $\ v^2=r g\ $ where $g$ is the
centripetal acceleration \cite{WEI}.
Dynamical observations show that the velocity of objects away
from the disk of the galaxy remains constant for large
distances, suggesting that there is dark matter with
$M(r)\propto r\ $  \cite{KT,RST}.

The analysis of the post--Newtonian motion of the halo can be
better described in standard coordinates \cite{WEI}
\begin{equation}
\label{DS2}
ds^2 = - B(r)dt^2 + A(r)dr^2 + r^2 d\Omega^2 \ \ .
\end{equation}
We have calculated the solutions to the equations of motion
(\ref{SEQ1}, \ref{SEQ2}) in the interior of the halo
($r<R_{\rm halo}$) of a spiral galaxy with mass
$M(r)=M_{\rm halo}{\displaystyle  \frac{r}{R_{\rm halo}} }$
and negligible pressure
\begin{equation}
\label{AGM}
\begin{array}{l}
B(r) =
{\displaystyle
1 + \frac{2GM(r)}{r}\ \ln\left(\frac{e^{-1}\ r}{R_{\rm halo}}\right)
+ ... }  \vspace{2mm}\\
A(r) =
{\displaystyle
1 - \left(\frac{\omega+1+\sigma}{\omega+2-\sigma}\right)
\frac{2GM(r)}{r}\ \ln\left(\frac{e^{-1}\ r}{R_{\rm halo}}\right)
+ ... } \vspace{3mm}\\
\Phi(r) \simeq
{\displaystyle
\left(\frac{e^{-1}\ r}{R_{\rm halo}}\right)^{\displaystyle
\frac{1-2\sigma}{\omega+2-\sigma}\ \frac{G M_{\rm halo}}
{R_{\rm halo}}} } \ \ ,
\end{array}
\end{equation}
continuosly connected with the post--Newtonian solution for
$r\geq R_{\rm halo}$ \cite{CQG}. The interior solutions
(\ref{AGM}) give a constant centrifugal velocity
\begin{equation}
\label{3KL}
v^2 \simeq\ \frac{GM(r)}{r} =\ \frac{G M_{\rm halo}}
{R_{\rm halo}} \ .
\end{equation}
Dark matter models of galaxy formation give generically a mass
distribution $\rho(r)=\rho_o r^n$ \cite{KT}. If the halo
were composed of particles with variable mass we would have
$\rho(r)\Phi^\sigma(r) = {\displaystyle
\frac{M_{\rm halo}}{4\pi R_{\rm halo}}\ \frac{1}{r^2} }$,
which corresponds to $M(r) \propto r$. From the solutions in the
interior of the halo (\ref{AGM}) we obtain the relation
\begin{equation}
\label{ALF}
n\ \simeq\ -2 + \frac{4\beta_I(\beta_I-\beta_V)}
{1+4\beta_I\beta_V} \times 10^{-6} \ \frac{G M_{\rm halo}}
{R_{\rm halo}} \ .
\end{equation}
Therefore, small deviations from $r^{-2}$ in $\rho(r)$ can be
accounted for by variable masses for dark matter. At scales of
spiral galaxies this effect is negligible but at larger scales,
say superclusters, it could change the general picture of dark
matter halos.

\section{Dark entropy and axion dilution}

Let us consider a perfect fluid composed of dark matter particles
with variable masses. We can write the energy--momentum tensor
and particle number current in the presence of a gravitational
field as $\ T^{\mu\nu} = p g^{\mu\nu} + (p+\rho) U^{\mu}U^{\nu}\
$ and $\ N^\mu = n U^\mu$, where $\ U^{\mu}\ $ is the local
value of ${\ \displaystyle \frac{dx^{\mu}}{d\tau}\ }$ for a
comoving fluid element and $n$ is the particle number density.

The general cosmological solutions to the equations of motion
(\ref{SEQ1}, \ref{SEQ2}) in a Robertson--Walker frame,
$ds^2 = - dt^2 + a^2(t)dx^2$
(compatible with the properties of a perfect fluid)
are \cite{NPB,CQG}
\begin{equation}
\label{SOL}
\begin{array}{l}
a(t)\sim t^{\ p}, \hspace{1cm}
{\displaystyle
p = \frac{2(2\omega+3)-2(1-2\sigma)}
{3(2\omega+3)-2(1-2\sigma)+(1-2\sigma)^2}
} \vspace{2mm} \\
\Phi(t)\sim t^{\ q}, \hspace{1cm}
{\displaystyle
q = \frac{4(1-2\sigma)}
{3(2\omega+3)-2(1-2\sigma)+(1-2\sigma)^2}  } \ \ ,
\vspace{1mm}
\end{array}
\end{equation}
while the energy and particle number conservation laws
expressed in a Robertson--Walker frame give
\begin{equation}
\label{CM}
\frac{d}{dt}(\rho a^3) + p \frac{d}{dt}(a^3) =
\frac{1}{m}\frac{d m}{dt} (\rho-3p) a^3
\end{equation}
\begin{equation}
\label{CN}
\frac{d}{dt}(n a^3) = 0 \ \ .
\end{equation}

Non--relativistic particles in thermal equilibrium at a
temperature $T$ have negligible pressure and energy density
given by
\begin{equation}
\label{RHO}
\rho = n\ m = \frac{N m}{a^3},
\end{equation}
where $N=na^3$ is the conserved number of dark matter particles
in thermal equilibrium. During the matter era there is an
entropy increase (for variable masses) that can be computed by
comparing eq.(\ref{CM}) with the second law of Thermodynamics
$dU + pdV = TdS$,
\begin{equation}
\label{TdS}
TdS \simeq\ N dm(\Phi) \ \ .
\end{equation}
The total entropy production per comoving volume from the time
of equal matter and radiation energy density to now, due to the
variable masses of dark matter, is given by
\begin{equation}
\label{TS}
\Delta S = \int_{t_{eq}}^{t_o} \frac{N dm}{T} \simeq
\frac{\sigma(1-2\sigma)}{2\omega+3 - (1-2\sigma)^2}\
\frac{N m(t_o)}{T_o} \equiv \ k(\omega,\sigma)\
\frac{N m}{T_o} \ \ ,
\end{equation}
where $T_o$ is the dark matter temperature
(approximately equal to the photon temperature) today
and we have used the cosmological solutions (\ref{SOL}).
This should be considered as a new source of entropy,
apart from the usual ones (cosmological phase transitions,
galaxy formation, star evolution, etc.).

We now apply this entropy production mechanism to the dilution
of axions \cite{PRL}, thus opening the so--called
``axion window" \footnote{For a detailed discussion of axions in
the early universe see Ref.\cite{KT}.}. Axions are very good
candidates for the dark matter of the universe \cite{RST}. They
are non--relativistic particles that couple very weakly to matter
and radiation. Astrophysical and cosmological constraints bound
their mass to be in the range $10^{-3}$--$10^{-5}$ eV $\lsim m_a
\lsim 10^{-2}$--$10^{-3}$ eV \cite{KT}. The lower bounds come
from the cosmological constraint $\Omega_a h^2 \lsim 1$, where
the axion contribution to the critical density is estimated
as \cite{KT}
\begin{equation}
\label{OMA}
\Omega_a h^2 \simeq\
\left(\frac{m_a}{10^{-3}-10^{-5}\ {\rm eV}}\right)^{-1.18}.
\end{equation}
and $h$ is Hubble's constant in units of 100 km s$^{-1}$
Mpc$^{-1}$ \cite{KT}. In the above estimation it was assumed
that there has been no significant entropy production at
later stages of the evolution of the universe. If, on the other
hand, the entropy per comoving volume $S$ is increased by a factor
$\gamma$ since the time of axion production, then $\Omega_a h^2$
is reduced by the same factor \cite{KT}. There has been several
attempts to open the ``axion window", the most important one
being the use of inflation \cite{INF}. Here we present an
alternative mechanism for axion dilution.

If axions had variable masses their energy density would be
diluted by a factor $\gamma=\frac{\Delta S}{S}$. Using the fact
that baryons are also non--relativistic, $\rho_B = n_B\ m_B$,
and their contribution to the critical density is $\Omega_B h^2
\sim 10^{-2}$ \cite{KT} we find, see eq.(\ref{TS}),
\begin{equation}
\label{RDS}
\gamma \simeq 10^2\ k(\omega,\sigma)\ \Omega_a h^2\
\frac{\eta N_\gamma}{S}\ \frac{m_B}{T_o},
\end{equation}
where $\eta=n_B/n_\gamma \sim 4\times 10^{-10}$ is the baryon
to photon ratio, and the total entropy per comoving volume of the
universe is related to the number of photons by $S\simeq 7
N_\gamma$ \cite{KT}. Note that the entropy production is
precisely proportional to the axion energy density (\ref{OMA})
and thus to the mass of the axion. Therefore, the fraction of
critical density contributed by axions with variable masses
does not depend on $m_a$,
\begin{equation}
\label{CRD}
\Omega_a h^2_{\ {\rm now}} \simeq \frac{\Omega_a h^2}{\gamma}
\simeq 7\times 10^{-2}\ k^{-1}(\omega,\sigma)\
\frac{T_o}{\eta\ m_B}.
\end{equation}
Imposing the bound $\Omega_a h^2_{\ {\rm now}} \lsim 1$,
we find {\em no} constraint on the axion mass but only
on the parameters of our model (\ref{SMD})
\begin{equation}
\label{PAR}
\frac{2\beta_I\beta_V - 2\beta_I^2}
{1-4\beta_I^2} \gsim 7\times 10^{-2}\ \frac{T_o}{\eta\ m_B},
\end{equation}
where $m_B\sim 1$ GeV is the proton mass and $T_o\sim 3$ K is
the photon temperature today. Variable masses of axions
could be an important alternative mechanism to inflation as a
source of axion dilution. It is important to know all possible
sources of axion dilution since there is a proposal of an
experimental search for dark matter axions \cite{SIK} in the
0.6--26 $\mu$eV mass range that would not detect an axion with
very small mass, on the other hand allowed by these processes.

\section{Experimental bounds}

As we have seen, the model has very original physical features
but we must constrain its parameters in order to appreciate its
quantitative relevance. Most of the bounds are cosmological.
Damour {\em et al.} \cite{DGG} gave bounds on the parameters
of the action (\ref{SMD}) from
radar time--delay measurements, the age of the universe and the
time variation of $G$. Visible baryonic matter dominates our
solar system and therefore the $\omega$ parameter
can be constrained by post--Newtonian experiments. In
particular, the Viking experiments \cite{VIK} give the bound
$\ 2\omega+3 > 500$ \hspace{1.5mm} (95\% c.l.), which imply, see
eq.(\ref{OSB}),
\begin{equation}
\label{BV2}
\beta_V < 0.022 \ \ .
\end{equation}
Cosmological observations give $\ H_o t_o > 0.4\ $ \cite{WAF},
where $H_o$ is the Hubble constant and $t_o$ is the age of the
Universe. This bound imposes a constraint on $\beta_I$ which is
almost independent of $\beta_V$, see eqs.(\ref{SOL}, \ref{OSB},
\ref{BV2}), and gives (see also \cite{DGG})
\begin{equation}
\label{BI2}
\beta_I < 0.674 \ \ .
\end{equation}

Since we are assuming that dark matter dominates the
cosmological evolution during the matter era, it will also
be constrained by bounds from primordial nucleosynthesis
\cite{DNS}. A recent analysis of these bounds was made in
Ref.\cite{PNS} for an ordinary Jordan--Brans--Dicke theory. We
have generalized
this result to a theory with $\Phi$--dependent masses, see also
Ref.\cite{DG}, by using the cosmological solutions (\ref{SOL})
and the fact that ${\displaystyle \ G\sim \frac{1}{\Phi}\ }$ and
\ $aT\sim$ constant.

The predicted mass fraction \ $Y_p$\ of primordial \ $^4{\em
He}$\ in this theory is correctly parametrized in the allowed
region of observable parameters by \cite{KAO,PNS}
\begin{equation}
\label{YP}
Y_p = 0.228 + 0.010\ln\eta_{10} + 0.012 (N_{\nu}-3) + 0.185
\left(\frac{\tau_n-889.6}{889.6}\right) + 0.327 \log \xi
\end{equation}
where $\eta_{10}$ is the baryon to photon ratio in units of
$10^{-10}$, $N_\nu$ the number of light neutrino species at
nucleosynthesis and $\tau_n$ is the neutron life--time
$\tau_n=889.6\pm 5.8 \ s \ \ (2\sigma)$ \cite{WMA}.
$\xi$ is the ratio of the Hubble parameter at nucleosynthesis to
its general relativity value and can be shown to be given in
Jordan--Brans--Dicke theory with
variable masses by, see eq.(\ref{SOL}),
\begin{equation}
\label{XG}
\xi^2\equiv\frac{G_{rad}}{G_o}
= \left(\frac{T_{eq}}{T_o}\right)^{\displaystyle \frac{q}{p}}
= \left( 2\times 10^4 \ \Omega_o h^2
\right)^{\displaystyle \frac{1-2\sigma}{\omega+1+\sigma}} \ \ ,
\end{equation}
where $\Omega_o$ is the observed ratio of the total energy
density of the universe to the critical density.
Present observations allow the range
$0.008 < \Omega_o h^2 < 4.0$ \cite{PDG}.

Consistency, within two standard deviations, of the
observational bounds on the primordial abundances of
$D +\ ^3{\em He}$ and $^7{\em Li}$  and the corresponding
predictions of GR for $N_\nu=3$, forces $\eta_{10}$ to lie in
the range $2.8 \leq \eta_{10} \leq 4.0$ \cite{KAO}. The lower
bound on $\eta_{10}$ comes from the upper observational limit on
$D +\ ^3{\em He}$ and is the relevant one for our purposes.

Using the conservative $2\sigma$ estimation of the observational
value of $Y_p$ \cite{KAO,MRE} (see however Ref.\cite{FBK})
\begin{equation}
\label{OBS}
Y_p = 0.23 \pm 0.01 \hspace{1cm} (2\sigma)
\end{equation}
we obtain a bound on $(\omega,\sigma)$ from primordial
nucleosynthesis, for $\Omega h^2=0.25$ and $N_\nu=3$
\cite{PNS,CQG}
\begin{equation}
\label{W380}
\frac{\omega+1+\sigma}{1-2\sigma}
>\ 380 \hspace{1cm} (95\%\ {\rm c.l.}) \ ,
\end{equation}
which constrains the parameters of our theory as
\begin{equation}
\label{BIV}
\beta_I\beta_V < 3\times 10^{-4} \ \ .
\end{equation}
Suppose that string theory or any other fundamental theory of gravity
fixes the coupling $\beta_I\sim {\cal O}(1)$. Then the constraint
(\ref{BIV}) would dramatically improve the bound on $\omega$ to
$2\omega+3 > 2\times 10^6$.

On the other hand, if axions had variable masses and constituted
the dark matter of our universe, their contribution to the
critical density would also impose a bound on the parameters of
our model, see eq.(\ref{PAR}). Using the previous bounds on
$\beta_I\beta_V$ (\ref{BIV}) we estimate
\begin{equation}
\label{CON}
\beta_I < 0.0165
\end{equation}
which improves significantly the bound on $\beta_I$, see
eq.(\ref{BI2}).

Furthermore, string theory predicts a scalar coupling of the
dilaton to the electromagnetic sector given by
$\alpha=\frac{e^2}{4\pi} \propto e^{\phi}$ (at tree level).
Therefore, a variation of the electromagnetic coupling constant
is related to the corresponding variation of Newton's constant by
\begin{equation}
\label{DAG}
\frac{\delta\alpha}{\alpha}=\sqrt{2\omega+3}\ \frac{\delta G}{G}.
\end{equation}
There are bounds on the variation of $\alpha$
which are extraordinarily strong \cite{DYS}.
For example, Dyson gives a bound
from the nuclear stability of the $\beta$--isotopes
$\ ^{187}_{75}Re$ and  $\ ^{187}_{76}Os$,
$\ |\frac{\delta\alpha}{\alpha}|<\ 2.5\times 10^{-5}\ $
since the formation of the Earth.
This bound translates into the parameters of our theory as
\begin{equation}
\label{STI}
\beta_I\beta_V<\ 1.7\times 10^{-8} \ \ .
\end{equation}
For a scenario in which all the masses have the {\em same}
$\Phi$--dependence ($\beta_I=\beta_V$), this imposes a
strong bound on $\omega$
\begin{equation}
\label{STO}
|2\omega+3| >\ 1.4\times 10^8 \ \ .
\end{equation}
As we can see, these bounds are much greater than any other
bound from nucleosynthesis or post--Newtonian experiments.
However, in string theory there are other scalar fields, the
moduli, that parametrize the size and shape of compactified
space and share many properties with the dilaton. In
particular, in some string models the gauge coupling is a
combination of the dilaton and moduli which could be fixed by
some mechanism, say gaugino condensation \cite{GAU}, without
fixing the dilaton. Therefore the last bound on the parameters
from the constancy of gauge couplings is a model dependent
constraint.

\section{Conclusions}

We have analyzed the phenomenology of a cosmological scenario in
which a dilaton scalar field couples differently to dark matter
than to visible matter. String theory gives no theoretical
prediction on the value of the dilaton coupling, it
just makes it plausible that this scenario may arise.
This kind of dilaton coupling violates the weak
equivalence principle but is not ruled out by experiment.

We study the physical consequences of such a simple asumption
and constrain the parameters of the model by experiment. Dark matter
particles with variable masses do not follow geodesics in the
conformal frame of visible matter with constant masses, and
therefore may have a measurable effect in the dynamical motion
of the halo of spiral galaxies. At scales of superclusters this
effect could change the general picture of dark matter halos.

As a consequence of variable masses, the energy--momentum tensor
of dark matter is not conserved. There is a ``dark entropy"
production associated with this dissipative effect. Axions are
very good candidates for the cold dark matter of our universe.
Their masses are constrained by astrophysics and cosmology to
lie in a very narrow range, the so--called ``axion window". If
axions had variable masses their contribution to the critical
energy density would be diluted and therefore open the axion
window. At the same time the cosmological constraint on the
axion energy density imposes a relatively strong bound on the
parameters of the model.

We assume that dark matter with variable masses dominates the
cosmological evolution during the matter era. The age of the
universe gives a bound on their dilaton coupling, while the
Viking experiment of radar time--delay bounds the coupling to
visible matter. If dark matter had variable masses, Newton's
constant would vary with time. In particular, it would be
different at the time of primordial nucleosynthesis.
Observational bounds on the mass fraction of primordial
$^4{\em He}$, $^3{\em He} + D$ and $^7{\em Li}$ also constrain
the parameters of the model.

In string theory the dilaton also couples to the electromagnetic
sector in the form of a variable gauge coupling. There are
bounds on the variation of the electromagnetic coupling that are
extraordinarily strong and therefore reduce significantly our
parameter space. However, as mentioned above, these bounds are
very model dependent. In some string scenarios the gauge
coupling is constant without fixing the dilaton.

Let us briefly summarize the cosmological bounds on the
parameters of the model, coming from radar time--delay
experiments, primordial nucleosynthesis and the contribution of
axions with variable masses to the critical density of the
universe
\begin{equation}
\label{COT}
\begin{array}{l}
\beta_V < 0.022 \\
\beta_I \beta_V < 3\times 10^{-4} \\
\beta_I < 0.017,
\end{array}
\end{equation}
which gives a rather small parameter space.

Finally, we think that this scenario for dark matter may be
interesting in the future, where models of structure formation
will have to take into account the recently observed anisotropy
of the cosmic background radiation. On the other hand, if dark
matter were composed of axions with variable masses, the lower
bound on their mass can be relaxed and may not be observed in
the recently proposed experimental search for axions.


\newpage
\begin{thebibliography}{99}
%
\bibitem{GSW} See, for instance, M. Green, J. Schwarz and E. Witten,
{\em Superstring Theory}  (Cambridge, 1987) and references therein.
%
\bibitem{CQG} J.A. Casas, J. Garc\'{\i}a-Bellido and M. Quir\'os,
{\em Class. Quant. Grav.} {\bf 9} (1992) (to appear).
%
\bibitem{DGG} T. Damour, G.W. Gibbons and C. Gundlach,
{\em Phys. Rev. Lett.} {\bf 64} (1990) 123.
%
\bibitem{JBD} P. Jordan, {\em Nature (London)} {\bf 164} (1949)
637; {\em Z. Phys.} {\bf 157} (1959) 112; C. Brans and
R.H. Dicke, {\em Phys. Rev.} {\bf 124} (1961) 925;
R.H. Dicke, {\em Phys. Rev.} {\bf 125} (1962) 2163;
C. Brans, {\em Phys. Rev.} {\bf 125} (1962) 2194.
%
\bibitem{SCT} For a general review, see C.M. Will,
{\em Theory and Experiment in Gravitational Physics}, Cambridge
University Press (1981); {\em Phys. Rep.} {\bf 113}
(1984) 345.
%
\bibitem{EI} R. Holman, E.W. Kolb and Y. Wang,
{\em Phys. Rev. Lett.} {\bf 65} (1990) 17;
J. Garc\'{\i}a-Bellido and M. Quir\'os,
{\em Nucl. Phys.} {\bf B 368} (1992) 463.
%
\bibitem{KT} E.W. Kolb and M.S. Turner, {\em The Early Universe},
Addison Wesley (1990).
%
\bibitem{NPB} J.A. Casas, J. Garc\'{\i}a-Bellido and M. Quir\'os,
{\em Nucl. Phys.} {\bf B361} (1991) 713.
%
\bibitem{WEI} See, for instance, S. Weinberg, {\em Gravitation and
Cosmology} (Wiley, New York, 1972).
%
\bibitem{RST} R. Sancisi and T.S. van Albada, in {\em Dark Matter in
the Universe}, eds. J.Kormady and G. Knapp (Reidel, Dordrecht, 1987).
%
\bibitem{PRL} J. Garc\'{\i}a-Bellido, {\em Axions with variable
masses}, preprint IEM--FT--55/92, April 1992.
%
\bibitem{INF} A.D. Linde, Phys.\ Lett.\ {\bf 201 B}, 437 (1988);
S. Dimopoulos and L.J. Hall, Phys.\ Rev.\ Lett.\ {\bf 60}, 1899
(1988); R.L. Davis, E.P.S. Shellard, Nucl.\ Phys.\ {\bf B 324},
167 (1989); M.S. Turner, F. Wilczek, Phys.\ Rev.\ Lett.\ {\bf
66}, 5 (1991); A.D. Linde, Phys.\ Lett.\ {\bf B 259}, 38 (1991).
%
\bibitem{SIK} P. Sikivie {\em et al.}, proposal to the DOE and
LLNL for an experimental search for dark matter axions in the
0.6--26 $\mu$eV mass range, IEEE Nuclear Science Symposium,
Santa Fe, 1991.
%
\bibitem{VIK} R.D. Reasenberg {\em et al.}, {\em Astrophys. J.
Lett.} {\bf 234} (1979) L219.
%
\bibitem{WAF} W.A. Fowler, {\em Q. J. Roy. Astron. Soc.}
{\bf 28} (1987) 87.
%
\bibitem{DNS} G. Steigman, {\em Nature} {\bf 261} (1976) 479;
D.N. Schramm and R.V. Wagoner, {\em Ann. Rev. Nucl. Sci.}
{\bf 27} (1977) 37; J.D. Barrow, {\em Mon. Not. R. Astr. Soc.}
{\bf 184} (1978) 677; J. Yang, D.N. Schramm, G. Steigman and
R.T. Rood, {\em Ap. J.} {\bf 227} (1979) 697.
%
\bibitem{PNS} J.A. Casas, J.Garc\'{\i}a-Bellido and M. Quir\'os,
{\em Mod. Phys. Lett.} {\bf A7} (1992) 447;
{\em Phys. Lett.} {\bf B278} (1992) 94.
%
\bibitem{DG} T. Damour and C. Gundlach,
{\em Phys. Rev.} {\bf D43} (1991) 3873.
%
\bibitem{KAO} K.A. Olive, D.N. Schramm, G. Steigman and T.F. Walker,
{\em Phys. Lett.} {\bf B236} (1990) 454;
K.A. Olive, D.N. Schramm, D. Thomas and T.P. Walker,
{\em Phys. Lett.} {\bf B265} (1991) 239; {\em Astrophys. J.} (1991)
(to appear); see also K.A. Olive, talk given at the {\em
Electroweak Physics Beyond the Standard Model} Workshop, October
1991, Valencia, Spain.
%
\bibitem{WMA} W. Mampe {\em et al.}, {\em Phys. Rev. Lett.} {\bf 63}
(1989) 593; W. Mampe, {\em Phys. World} {\bf 4} (1990) 41.
%
\bibitem{PDG} Particle Data Group, Review of Particle Properties,
{\em Phys. Lett.} {\bf B238} (1990) 1.
%
\bibitem{MRE} M. Rees, {\em Nucl. Phys. B} (Proc. Suppl.)
{\bf 16} (1990) 3.
%
\bibitem{FBK} G.M. Fuller, R.N. Boyd and J.D. Kalen,
{\em Astrophys. J. Lett.} {\bf 371} (1991) L11.
%
\bibitem{DYS} F.J Dyson, {\em Phys. Rev. Lett.} {\bf 19} (1967) 1291;
J.N. Bahcall and M. Schmidt, {\em Phys. Rev. Lett.} {\bf 19} (1967) 1294;
A.M. Wolfe, R.L. Brown and M.S. Roberts,
{\em Phys. Rev. Lett.} {\bf 37} (1976) 179;
M. Maurette, {\em Annu. Rev. Nucl. Sci.} {\bf 26} (1976) 319;
A.I. Shlyakhter, {\em Nature} {\bf 264} (1976) 340; A.D. Tubbs and
A.M. Wolfe, {\em Astrophys. J. Lett.} {\bf 236} (1980) L105.
%
\bibitem{GAU} V.A. Novikov, M.A. Shifman, A.I. Vainshtein and
V.I. Zakharov, {\em Nucl. Phys.} {\bf B229} (1983) 407;
J.P. Derendinger, L.E. Ib\'a\~nez and H.P. Nilles,
{\em Phys. Lett.} {\bf 155B} (1985) 65;
M. Dine, R. Rohm, N. Seiberg and E. Witten,
{\em Phys. Lett.} {\bf 156B} (1985) 55.
%



\end{thebibliography}
\end{document}